\documentclass[review]{elsarticle}
\usepackage{amsmath,amssymb,amsfonts}
\usepackage{algorithmic}
\usepackage{graphicx}
\usepackage{textcomp}
\usepackage{xcolor}
\usepackage{textcomp}
\usepackage{listings}
\usepackage{color}
\usepackage{tikz}
\usepackage{url}
\usepackage{tikz-network}
\usepackage{subcaption}
\usepackage[inline]{enumitem}
\usepackage{tabularx}
\newcolumntype{Y}{>{\centering\arraybackslash}X}
\definecolor{dkgreen}{rgb}{0,0.6,0}
\definecolor{gray}{rgb}{0.5,0.5,0.5}
\definecolor{mauve}{rgb}{0.58,0,0.82}

\def\BibTeX{{\rm B\kern-.05em{\sc i\kern-.025em b}\kern-.08em
    T\kern-.1667em\lower.7ex\hbox{E}\kern-.125emX}}

\journal{Expert Systems with Applications}

\begin{document}
\begin{frontmatter}
\title{Cross Version Defect Prediction with Class Dependency Embeddings}
\author[add1]{Moti Cohen}
\ead{mordechc@post.bgu.ac.il}
\ead{cohmoti@gmail.com}
\author[add1]{Lior Rokach}
\ead{liorrk@post.bgu.ac.il}
\author[add1]{Rami Puzis}
\ead{puzis@bgu.ac.il}

\address[add1]{	Department of Software and Information Systems Engineering, Ben Gurion University of The Negev}

\begin{abstract}
Software Defect Prediction aims at predicting which software modules are the most probable to contain defects. The idea behind this approach is to save time during the development process by helping find bugs early. Defect Prediction models are based on historical data. Specifically, one can use data collected from past software distributions, or \textit{Versions}, of the same target application under analysis. Defect Prediction based on past versions is called Cross Version Defect Prediction (CVDP). Traditionally, Static Code Metrics are used to predict defects. In this work, we use the Class Dependency Network (CDN) as another predictor for defects, combined with static code metrics. CDN data contains structural information about the target application being analyzed. Usually, CDN data is analyzed using different handcrafted network measures, like Social Network metrics. Our approach uses network embedding techniques to leverage CDN information without having to build the metrics manually. In order to use the embeddings between versions, we incorporate different embedding alignment techniques. To evaluate our approach, we performed experiments on 24 software release pairs and compared it against several benchmark methods. In these experiments, we analyzed the performance of two different graph embedding techniques, three anchor selection approaches, and two alignment techniques. We also built a meta-model based on two different embeddings and achieved a statistically significant improvement in AUC of 4.7\% ($p < 0.002$) over the baseline method.
\end{abstract}

\end{frontmatter}
\newpage
\section{Introduction}
Software Development is a very long and complicated process, involving many stages and multiple participants. More so, today's software systems have become large and highly complex. Given these characteristics of the software development process, it is inevitable that some defects will end up in applications. Software Quality Assurance procedures provide a means to try and locate these defects. These procedures are very time-consuming and not complete, and usually involve intensive human intervention (unit test writing, for example). In order to help focus quality assurance efforts, many defect prediction and detection approaches were developed. In the last few decades, much progress has been made in using machine learning techniques to help in defect prediction and identification \cite{kamei_defect_2016-1}.

In the field of Defect Prediction, one can divide the different approaches into two main categories. The first is Cross Project Defect Prediction (\textbf{CPDP}), which performs transfer learning from one project to another. The main issue in \textit{CPDP} is to find from which project to learn and how to transfer the learned model so that the model will be valid for the target project.

The second category is Within Project Defect Prediction (\textbf{WPDP}) aims at predicting defective code in a given project using the same project's past data. This approach can be divided into two subcategories: Inner Version Defect Prediction (\textbf{IVDP}) and Cross Version Defect Prediction (\textbf{CVDP}). In \textbf{IVDP}, data from the same project version is used as training data and test data, while in \textbf{CVDP}, data from past project versions are used as training data while the latest ("new") version is the test data. The data in \textbf{IVDP} is usually more homogeneous, but this scenario is usually less realistic, and the data available is usually sparse. In the case of \textbf{CVDP}, there is usually more data, but the distribution tends to change between versions, making it hard to transfer the knowledge. The focus of our work is on the \textbf{CVDP} scenario.

Class Dependency data was used in different studies to predict software defects(e.g.\cite{zimmermann_predicting_2008-1,qu_node2defect_2018},\cite{premraj_network_2011},\cite{ma_empirical_2016}). Most studies use manually crafted features over the Class Dependency Network, usually Social Network measures. In \cite{qu_node2defect_2018}, graph embedding was first used to generate automatic features from CDN for the Defect Prediction process, specifically in a \textbf{IVDP} setup. In this work, we try to apply the same methodology but in the more complex \textbf{CVDP} setup. This higher complexity is because software modules exhibit different statistics in different versions\cite{xu_cross_2018}. In order to use embeddings in a cross-version setup, we need to align the embeddings learned in the test set to the embeddings from the train set. We use different alignment techniques and combine the aligned embeddings with traditional static code metrics to train a classifier and achieve an improvement over the state-of-the-art baseline method.

The main contributions of our study are: 
\begin{itemize}
    \item We develop a novel approach to module level CVDP, based on a combination of static code metrics and CDN embeddings. We incorporate the use of embedding alignment techniques when moving from one version to another.
    \item We developed two anchor selection techniques for the alignment process and evaluated them with two alignment procedures. We also used two different embedding frameworks during the experiments.
    \item We performed an experimental analysis of our techniques on a public dataset, with nine software projects written in Java\cite{noauthor_java_nodate}. These projects contain a total of 24 (old version, new version) pairs. We calculated and analyzed two performance metrics, AUC and F1-score, and compared them to a state-of-the-art baseline technique. 
    \item We built a meta-model that combines the best models for each of the embedding techniques. Our meta-model achieves an improvement of 4.7\% in the AUC score over the baseline method.
\end{itemize}
\begin{figure*}[ht]
\centering
\includegraphics[width=\linewidth]{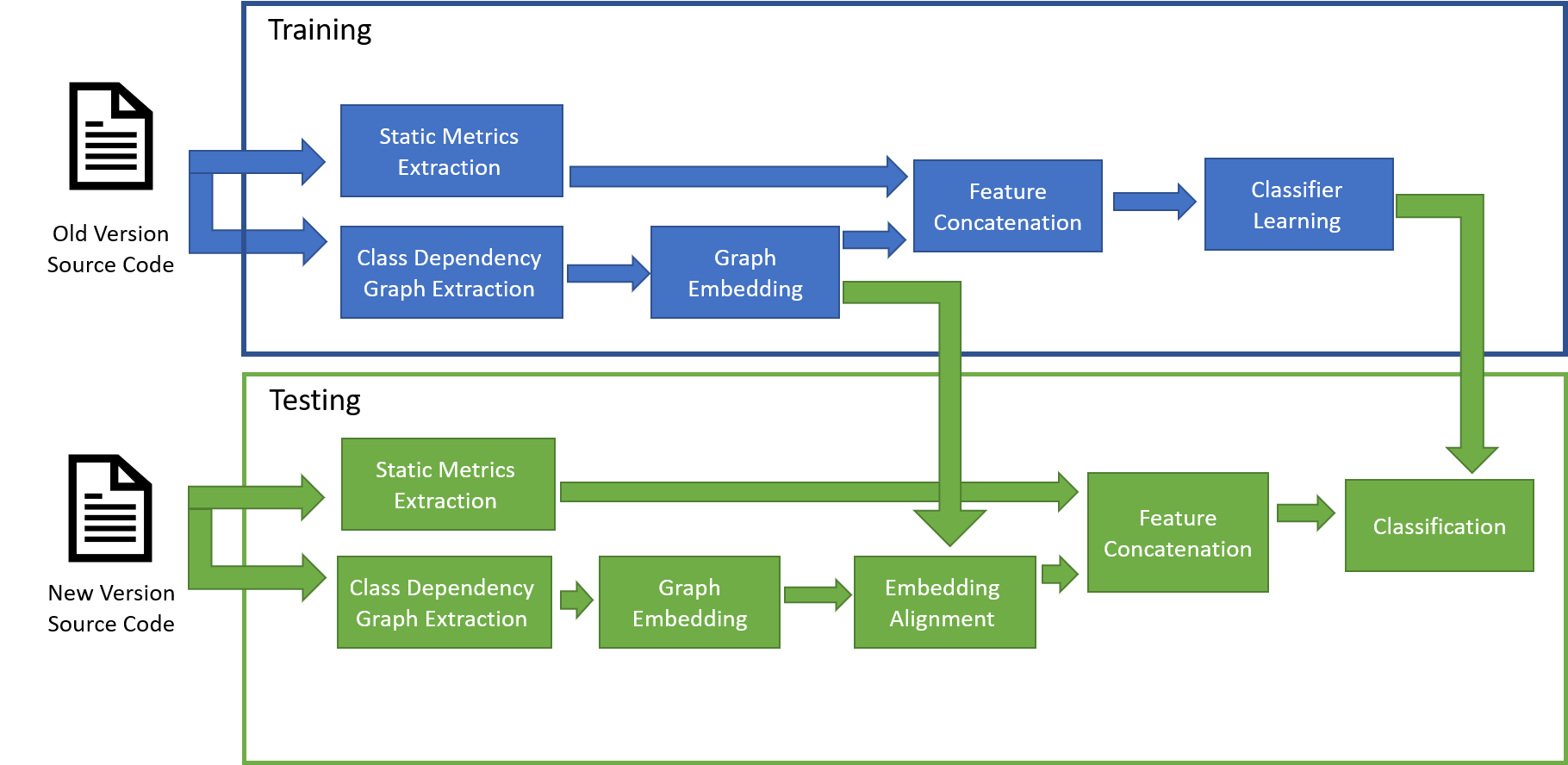}
\caption{A Flowchart of Our Proposed Solution }
\label{fig:flow}
\end{figure*}

\section{Related Work}
\subsection{Cross Version Defect Prediction}
There have been a lot of research done in the area of Defect Prediction \cite{nam_survey_2014, hall_systematic_2012,dambros_extensive_2010}. Most of these studies were done in the areas of same-version and cross-project defect prediction. To the best of our knowledge, the field of cross-version defect prediction has been less studied. Zimmerman et al. \cite{zimmermann_predicting_2007} were one of the first who showed that models learned from past versions data are useful in predicting defects in future releases. They conducted an experiment on Eclipse bug data over three releases and showed that models learned from past releases give consistent results compared to same-version models. 
In \cite{bennin_empirical_2016}, the authors conducted a cross-version evaluation of 11 different prediction models. The dataset used was 25 open source projects, having two releases each. They used seven process metrics as the features, and no code features. Bennin et al. \cite{bennin_significant_2017} analyzed the impact of data sampling on CVDP. They used 20 pairs of software versions and analyzed the impact of six data sampling techniques on classification performance. They also used five different classifiers for the experiment. They concluded that data sampling is beneficial for defect prediction, given that many datasets are imbalanced. Xu et al. \cite{xu_cross-version_2018} tried to tackle the issue that software applications evolve, and with it, the distribution of different metrics collected. To address this issue, they devised a two-phase framework for CVDP. The first step (called HALKP) is a hybrid active learning technique which selects modules from the current version (which are unlabeled), based on different measures, and consults an expert that assigns a label to it. This labeled instance is merged with the previous version's dataset, and the process continues. When they decide to stop the process, the result is a dataset of combined previous and current version instances. The second phase performs Kernel PCA (KPCA) to map all instances (labeled and unlabeled) to a new feature space. The result of this process is fed into a regular classifier. They show that their technique improves classification performance over a baseline with just the original features. Amasaki \cite{amasaki_cross-version_2018} investigated the performance of CPDP techniques in a CVDP scenario. In this study, Amasaki evaluated the performance of 20 CPDP techniques in two different scenarios: \textit{Single Older Version(SOV)} and \textit{Multiple Older Versions(MOV)}. They used 11 different projects for the experiment, with 3 to 5 versions each. The experiment had two interesting results: \begin{enumerate*}\item CPDP techniques can be useful in CVDP scenario\item Using multiple past versions also improves the prediction performance.\end{enumerate*} Yang et al.\cite{yang_ridge_2018} performed a different CVDP study, trying to sort software modules according to defect count prediction. The idea was to predict which modules might contain more bugs, rather than simply predicting if a module has or does not have at least one bug. They investigated the performance of Ridge and Lasso Regression in this sorting problem. They concluded that although the datasets used have some issues, the analyzed methods perform better than linear regression and negative binomial regression on CVDP. Shukla et al.\cite{shukla_multi-objective_2018} formulated the CVDP problem as a multi-objective optimization problem, trying to maximize recall by minimizing classification error and QA cost. They compare this method with basic models and show they have improved recall and achieved a smaller misclassification error. In \cite{xu_cross_2018} the authors tried to tackle the distribution dissimilarity between versions problem using a subset selection algorithm called \textit{Dissimilarity based Sparse Subset Selection(\textbf{DS3})}. The idea is to select a subset of modules from a past version that best represents the distribution of the current version, and use these modules as the training set. They compared this technique with simply using all the original data as the training set, and showed improved performance in classification accuracy and effort aware metrics. The authors of \cite{XU201959} tried to improve this technique further by adding a subset selection from the previous version, which represents the data well. This subset is later fed into \textbf{DS3}, similar to the previous work. They show improved results for regular and effort aware metrics. For this work, they use Static Metrics only. In \cite{fan_software_2019}, the authors used an Attention Based Recurrent Neural Network to encode AST information and evaluated the performance of this method in a CVDP scenario. They show that this technique achieves promising results.

\subsection{Class Dependency Network}
The idea behind CDN is to try and capture the structural information of a given software application. Capturing structural information is achieved by creating a graph with different relations between software components or modules. Formally, a CDN is defined \cite{subelj_community_2011} as a multi-graph $G=(N,E)$ where $N$ is the set of nodes (modules), and $E$ is the set of edges (relations).  

Class Dependency information has been used in different defect prediction studies. In their seminal work, Zimmerman et al. \cite{zimmermann_predicting_2008-1} studied defect prediction performance when using a combination of network measures and static code metrics. They concluded that network measures increase recall by 10\%, when performed classification on windows server defect data. Premraj et al.\cite{premraj_network_2011} replicated the experiment on different projects and in a CVDP setup and concluded that in a CVDP scenario, the network measures offered no added value. On the other hand, the authors of \cite{ma_empirical_2016} performed a different evaluation and concluded that network measures have a positive relation with defects, and can be used in defect prediction. They also pointed out that these effects are not consistent in some projects. In \cite{qu_node2defect_2018}, the authors used CDN embedding as the features in a same-version defect prediction scenario, combined with static code metrics, and showed that this could improve the results of defect prediction which uses only static code metrics. In a recent study, Qu et al. \cite{qu_using_2019} used a new approach to analyze CDN for defect prediction. They used K-core Decomposition on the CDN and observed that modules within high k-cores have a higher probability of being buggy. They used this new observation in both IVDP and CVDP scenarios and showed an improvement over baseline methods in an effort-aware bug prediction scenario.

\subsection{Graph Embedding}\label{GraphEmb}
Graph (or network) Embedding is the process of generating a representation of graph elements in a vector space, which preserves some property desired. There has been much research in the area\cite{goyal_graph_2017}, and many techniques and applications have been researched and proposed. Graph embedding techniques have been used in multiple domains and provided great results. One main advantage of these techniques is their ability to extract features from graph data automatically, usually with minor tuning. We will provide a short review of the two embedding techniques used in this work: \textit{Node2vec}\cite{grover_node2vec_2016} and \textit{LINE}\cite{tang_line_2015}.

\subsubsection{Node2vec}
\textit{Node2vec} is a framework for graph embedding. The basic idea comes from a Natural Language Processing model called Skip-gram \cite{mikolov_efficient_2013}. The idea in \textit{Node2vec} is to maximize the probability of observing a node's neighborhood, given its vector representation. The algorithm tries to learn a vector representation that tries to maximize that probability. A node's neighborhood is defined by a sampling strategy, meaning a node can have different neighborhoods.  

The \textit{Node2vec} algorithm works as follows. For each node in the graph, we sample a set of random walks from it. The length of the walk, the number of walks and the sampling strategy can be modified. After sampling the walks, each walk is used as a "sentence" in natural language, to represent a node's context. These walks are used in the learning process of the embedding function, which maps each node into a k dimensional embedding (also a parameter). The sampling strategy used in \textit{Node2vec} is based on a sampling rule the authors define as a $2^{nd}$ order random walk with two parameters: $p$ and $q$. Generally speaking, a sampling strategy can be biased towards "walking" farther from the start node (like a Depth First Search) or be biased towards "staying" close to the start node (like a Breadth-First Search). A high parameter $p$ value (relative to $q$) causes a more DFS like sampling, and a high $q$ value (relative to $p$) causes a more BFS like sampling.

\subsubsection{LINE}
The idea behind \textit{LINE} (\textbf{L}arge-scale \textbf{I}nformation \textbf{N}etwork \textbf{E}mbedding) is to generate an embedding that models node proximity similarity. We use second-order proximity because our graph is directed. Second-order proximity assumes that similar nodes have similar "neighborhoods", meaning connections to nodes. So the idea is to model these connections and to learn the node similarity based on these connections. Each node is modeled both as a node and as a "context" (like modeling its connections). Later, we measure the probability of getting a node given a context (a different node), and we look for a representation that generates a distribution as similar as possible to the empirical distribution, as defined in the original paper.  

\subsection {Embedding Alignment}
As described earlier, a graph embedding is a vector representation for each node or edge, to preserve some target measure. When describing embeddings, we did not constrain the resulting embedding in any way, meaning there can be many different embeddings which have the same "results". For example, in a Euclidean vector space, every rotation of an embedding will preserve the Euclidean distances in that embedding. Even running the same embedding algorithm on the same dataset can result in a very different embedding space. In case we have two embeddings of "similar" items, for example, embeddings of words from two languages, we might want to align those two embeddings to the same coordinates system as close as possible while preserving the relations between the data points. Performing the alignment can give us a unified representation of the embeddings, enabling operations between the datasets. These techniques have been used in Natural Language Processing\cite{smith_offline_2017, xing_normalized_2015}. We will describe an alignment procedure with parallel points bet1ween two embeddings.

We start with two embeddings, $E_{1}$ and $E_{2}$. Our goal is to build a linear transformation $T$ which maps from $E_{2}$ to $E_{1}$. We are also provided with a set of parallel points, $(x^{(i)},y^{(i)}), x^{(i)} \in E_{2}, y^{(i)} \in E_{1}$. We wish to build $T$ s.t. $T(x^{(i)}) \approx y^{(I)}$. Let $\textbf{X}$ and $\textbf{Y}$ be the matrices whose columns are vectors $x^{(i)}$ and $y^{(i)}$ respectively. Then we wish to solve
\begin{equation}\label{eq:pro}
\min_{T}\|Y-TX\|_{F}    
\end{equation}
where $\|A\| = \sqrt{\sum_{i,j}{{|a_{ij}|}^2}}$ is the Frobenius norm. The general problem is hard to solve, but in constraining the solution to be orthogonal matrices, we get the Orthogonal Procrustes problem, which has a closed-form solution. An orthogonal matrix $Q$ is defined by $Q^{T}Q = QQ^{T} = I$, where $I$ is the identity matrix. Schönemann \cite{schonemann_generalized_1966} found the closed-form solution. if $U\Sigma V$ is the SVD decomposition of $YX^{T}$, then the solution to \eqref{eq:pro} is given by $T = UV^{T}$. To the best of our knowledge, no one has tried using this technique in code embedding alignment.

\section{Proposed Framework for CVDP}
In this section, we will describe our solution framework for CVDP.

\subsection{Framework Overview}
Given a pair of software versions $(\mathcal{V}_{0}, \mathcal{V}_{1})$, where $\mathcal{V}_{0}$ is a prior version to $\mathcal{V}_{1}$, we would like to build a defect classifier for the modules in $\mathcal{V}_{1}$. Our solution is composed of a few steps. In the training phase, we do the following:

\begin{enumerate}
    \item Calculate Static Code Metrics for $\mathcal{V}_{0}$ [Section \ref{StaticCodeMetrics}]
    \item Extract CDN for $\mathcal{V}_{0}$, marked as $G_{0}=(V_{0},E_{0})$  [Section \ref{CDNE}].
    \item Learn an embedding for $G_{0}$ [Section \ref{EmbedLearn}].
    \item Learn a Classifier using all the data available for $\mathcal{V}_{0}$ [Section \ref{ClsLearn}].
\end{enumerate}

After we built our training model, these are the steps for the classification phase:
\begin{enumerate}
    \item Calculate Static Code Metrics for $\mathcal{V}_{1}$.
    \item Extract CDN for $\mathcal{V}_{1}$, marked as $G_{1}=(V_{1},E_{1})$.
    \item Learn an embedding for $G_{1}$.
    \item Perform embedding alignment between the embedding for $\mathcal{V}_{1}$ and the embedding for $V_{0}$ [Section \ref{EmbedAlign}].
    \item Perform classification using the aligned embedding for $\mathcal{V}_{1}$.
\end{enumerate}
In the following sections we will describe in detail the different steps of our framework, and discuss different considerations that arose during the study.

\subsection{Static Code Metrics} \label{StaticCodeMetrics}
Static Code Metrics are the classic and state-of-the-art metrics used in defect prediction. These metrics exist for decades and many different metrics were developed during the years\cite{fenton_software_2014}. Most of these metrics try to capture code complexity and size, bad class design etc. We use the metrics defined by \cite{chidamber_metrics_1994,henderson-sellers_object-oriented_1996, bansiya_hierarchical_2002,martin_oo_1994,tang_empirical_1999,mccabe_complexity_1976}. The metrics used are described in Table \ref{table:2}. 

\begin{table}[ht]

    \centering
    \begin{tabular}{||c||}
         \hline
        Metric Name \\
        \hline\hline
        Weighted methods per class\cite{chidamber_metrics_1994}\\
        \hline
        Depth of Inheritance Tree\cite{chidamber_metrics_1994}\\
        \hline
        Number of Children\cite{chidamber_metrics_1994}\\
        \hline
        Coupling Between Object classes\cite{chidamber_metrics_1994}\\
        \hline
         Response for a Class\cite{chidamber_metrics_1994}\\
        \hline
        Lack of cohesion in methods\cite{chidamber_metrics_1994}\\
        \hline
        Lack of cohesion in methods 3\cite{henderson-sellers_object-oriented_1996}\\
        \hline
        Number of Public Methods\cite{bansiya_hierarchical_2002}\\
        \hline
        Data Access Metric\cite{bansiya_hierarchical_2002}\\
        \hline
        Measure of Aggregation \cite{bansiya_hierarchical_2002}\\
        \hline
        Measure of Functional Abstraction\cite{bansiya_hierarchical_2002}\\
        \hline
        Cohesion Among Methods of Class\cite{bansiya_hierarchical_2002}\\
        \hline
        Inheritance Coupling \cite{tang_empirical_1999}\\
        \hline
        Coupling Between Methods\cite{tang_empirical_1999}\\
        \hline
        Average Method Complexity\cite{tang_empirical_1999}\\
        \hline
        Afferent Couplings\cite{martin_oo_1994}\\
        \hline
        Efferent Couplings\cite{martin_oo_1994}\\
        \hline
        Average McCabe's Cyclomatic Complexity\cite{mccabe_complexity_1976}\\
        \hline
        Maximum McCabe's Cyclomatic Complexity\cite{mccabe_complexity_1976}\\
        \hline
        Lines of Code\\
        \hline
    \end{tabular}
    \caption{Static Code Metrics list}
    \label{table:2}
\end{table}

\subsection{Class Dependency Network Extraction}\label{CDNE}
We have defined CDN formally in a previous section. We slightly modify this definition and define the CDN to be directed, unlike in \cite{subelj_community_2011}. Each Edge $e_{i} \in E$ has a type associated with it, and the set of edge types defined by $T$.

Since we are analyzing Java programs, our components are classes, interfaces, annotations, and enumerations. We extracted a total of 10 relation types($T$), described in table \ref{table:1}. These edge types are based on the interactions between types in the Java programming language. This list contains most relation types. We chose not to handle relations based on Generic types since these were not very common in our dataset. In Figure 1 there is an example java code with different software dependencies, and the CDN generated from it. This example demonstrates a subset of our recognized types. We have written a tool that parses Java source code and builds the CDN based on references in the source code, and not the compiled version of the application since some changes can occur due to compiler optimizations. The resulting artifact is a single graph for each software version, containing the nodes and edges as described above.

The CDN extraction process runs in linear time and is composed of two passes. The first pass parses all Java files in a project repository and constructs a type dictionary for the project. The second pass traverses the ASTs of the code, analyzes the statements and extracts type references. These references are looked up in the dictionary built in the first pass. The extracted relations are appended to the CDN.`

\begin{table}[ht]
    \centering
    \resizebox{\columnwidth}{!}{
    \begin{tabular}{||c|c||}
         \hline
        Edge Type & Description \\ [0.5ex] 
        \hline\hline
        Extends[E] & A class extends another class\\
        \hline
        Implements[I] & A class implements an interface\\
        \hline
        Return Type[R] & A Type appears as a return type in a function\\
        \hline
        Variable[V] & A Type appears as a variable type in a function\\
        \hline
        Class Member[CM] & A Type appears as a class member type\\
        \hline
        Object Instantiation[OI] & A Type appears in a "new" statement\\
        \hline
        Annotation[A] & A Type appears as an annotation\\
        \hline
        Parameter[P] & A Type appears as a parameter type in a function\\
        \hline
        Static Class Member[SCM] & A Type appears as a static class member type\\
        \hline
        Static Method Call[SMC] & A class calls a static method from another class\\
        \hline
    \end{tabular}}
    \caption{List of edge types in our CDN}
    \label{table:1}
\end{table}

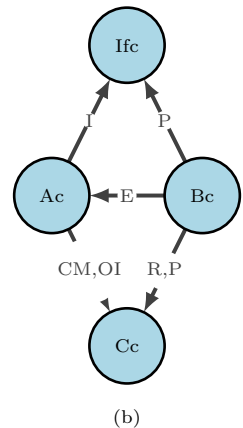
\begin{figure}
\centering
\begin{subfigure}{.5\linewidth}
  \centering
\begin{lstlisting}[language=java]
interface Ifc{
    void f();
}

public class Ac implements Ifc{
    private Cc c;
    
    public void f(){
        c = new Cc();
    }
}

public class Bc extends Ac{
    public Cc f2(Ifc i, Cc c2){
        i.f();
        return this.c;
    }
}

public class Cc{
//...
}
\end{lstlisting}  
  \caption{}
  \label{fig:sub1}
\end{subfigure}%
\begin{subfigure}{.5\linewidth}
  \centering
  \begin{tikzpicture}
  
  \Vertex[y=2,x=0,label=Ac,size=1]{Ac}
  \Vertex[y=4,x=1,label=Ifc,size=1]{Ifc}
  \Vertex[y=2,x=2,label=Bc,size=1]{Bc}
  \Vertex[y=0,x=1,label=Cc,size=1]{Cc}
  \Edge[Direct,label=I](Ac)(Ifc)
  \Edge[Direct,label={CM,OI}](Ac)(Cc)
  \Edge[Direct,label={E}](Bc)(Ac)
  \Edge[Direct,label={R,P}](Bc)(Cc)
  \Edge[Direct,label={P}](Bc)(Ifc)

\end{tikzpicture}

  \caption{}
  \label{fig:sub2}
\end{subfigure}
\caption{Code and extracted CDN example }
\label{fig:test}
\end{figure}
\subsection{CDN Embedding Learning}\label{EmbedLearn}
As described earlier, we use a few different graph embedding algorithms for the embedding process. We use the CDN extracted from the source code and generate a stripped graph that is directed and without types. The stripping means that if two types have multi connections in the CDN (in the same direction), in the stripped graph they will have a single directed connection. In case two types point to each other, there will be two edges. We do not give weights to the edges. Some classes do not exist in the CDN, so they will not have an embedding. For the process of embedding generation, we use the algorithms described in Section \ref{GraphEmb}.

\subsection{Embedding Alignment}\label{EmbedAlign}
As we discussed, the key to performing CVDP is to align the two version's embeddings, so that we will have a close as possible coordinates system. For this, we used an Orthogonal transformation to map between the embeddings. We also experimented with Linear Regression as a benchmark alignment technique. The relevant results will be discussed in section\ref{ExpRes}. The reason we chose to use an Orthogonal Transformation is that these transformations preserve angles between vectors and vector lengths. Because of these properties, vector distances (euclidean) are preserved, and hence the relations between the embedded elements. Intuitively, an Orthogonal Transformation does not distort the embedding but rotates and reflects it.

An essential part of the alignment procedure is to select the parallel points (or anchors) correctly. Poorly selected anchors can degrade results. In our setting, the nodes are software elements. Since we analyze a pair of versions of the same application, we expect most elements from the old version to be present in the newer version. Our goal is to select a subset of these types as our anchors. To do this, we used two techniques and compared them. For each technique, we calculate a score for a given node and select the $\mathcal{N}$ anchors with the highest scores. We performed experiments with different $\mathcal{N}$ values and the results will be discussed in Section \ref{ExpRes}.

\subsubsection{K-Nearest Neighbors Anchor Selection}
The motivation behind embedding is to generate a vector representation that preserves semantic relations. So, we expect nodes with similar structures in the graph to be located relatively close in an embedding space. We also assume that a node's close neighborhood has semantic meaning. Our assumption is as follows: given a node that exists in both graphs, we assume that if its structural behavior did not change between the graphs, its neighborhood would not change as well. This means that a node with high similarity in its neighbors group should get a high score. Formally, for each node $v_{i}\in V_{0}\cap V_{1}$ we calculate the following KNN score:
$$S_{knn}(v_{i}) = \frac{|N^i_{0}\cap N^i_{1}|}{k}$$
Where $N^i_{0}$ and $N^i_{1}$ are node i's $k$ nearest neighbors in $G_{0}$ and $G_{1}$ respectively. As this ratio gets closer to 1, the greater the similarity between the versions for that specific node. Each neighborhood $N^i_{j}$ is the set of closest nodes in the respective graph's embedding space, using the euclidean distance metric. We have experimented with other metrics, such as cosine similarity, and have achieved similar results.

\subsubsection{Graph Neighbors Similarity Anchor Selection}
The idea behind this technique is similar to the prior one, but from the original graph's point of view. Given a node that exists in both graphs, we extract its direct neighbors in each graph. We then look at the intersection of those groups, and reward nodes with a large intersection. We also assumed that nodes with a high degree and a high similarity can be more important, and the experiment showed that. Formally, we define this measure as follows:
$$S_{GNS}(v_{i}) = \frac{|M^i_{0} \cap M^i_{1}|^2}{|M^i_{0} \cup M^i_{1}|}$$
Where $M^i_{0}$ and $M^i_{1}$ are node i's immediate neighbors group in $G_{0}$ and $G_{1}$ respectively. Formally,
$$M^i_{j} = \{v_{l} | (v_{i},v_{l}) \in E_{j}  \vee  (v_{l},v_{i}) \in E_{j}\}$$
Where
$$v_{l},v_{i} \in V_{j} , j \in \{0,1\}, i,l \in \{1, ... ,|V_{j}|\}$$

\subsection{Classifier Learning}\label{ClsLearn}
In the previous sections, we demonstrated how we perform CDN extraction, embedding learning, and alignment. We also mentioned the use of static code metrics as additional features. The way we use both feature sets is rather straight forward. We concatenate both the static code metrics and the embedding values into a unified set and use this new set as our training/test data, which is fed into a regular classifier. The classification goal is to predict if a software module contains a bug or not. In the experiments, a Random Forest\cite{breiman_random_2001} classifier was used. Because Random Forest is based on multiple random decisions, each experiment we performed was repeated 30 times to generate an average estimate that is less biased by randomness. We chose the Random Forest classifier because of its popularity in Defect Prediction setups, and because it showed promising results in our experiments.

\section{Experimental Setup}\label{ExpSetup}
To evaluate our methods, we experimented with real-world software applications, comparing our results with two baselines. First, we build a classifier with only static code metrics as the features.  Second, we want to build baseline techniques that use embeddings and alignments. For this purpose we provide two models. The first model uses the learned embeddings without performing alignment. This shows the need for performing some alignment. The second model uses Linear Regression as a benchmark alignment technique. 

The Linear Regression alignment is rather straightforward. Linear Regression learns a linear relation between a set of variables and a target variable. In our setup, we wish to represent the new version's embeddings as a relation of the old one's embeddings. As was discussed earlier, the idea is to learn a mapping $T$ between  
$E_{1}$ and $E_{2}$. This can be broken down to $k$ different linear relations (n is the embedding size). So, given our anchors set, we build k linear regression problems from $E_{1}$ to each of the dimensions of $E_{2}$. These regression problems have a zero coefficient for simplicity. Our embedding alignment matrix $T$ is composed of the learned coefficients.

In our experiments, we used static code metrics collected by Jureczko et al.\cite{jureczko_towards_2010}. Not all projects that are reported in this dataset have an available source code, so we used only the ones we could locate the relevant sources. Also, only projects with more than one version available were used. We used data from a total of 9 projects, and a total of 24 version pairs. Table \ref{table:3} describes the different projects and versions analyzed in the experiments. Version pairs were chosen based on the dataset, where consecutive versions were paired. The original dataset does not cover all versions, so version jumps are sometimes significant. We collected the source code of each version from the relevant project's website, including all peripheral code (tests, for example). This code is used during our CDN construction and provides additional knowledge on the structural dependencies in the core application code. During embedding generation, some software modules do not have an embedding, due to a lack of graph edges in the CDN. To make a fair analysis, we only keep modules that have both static code metrics and an embedding. In table\ref{table:4} we provide a summary of the dataset's statistics. For each project, we calculate average measures of CDN size (vertices and edges), the number of modules that have both an embedding and static code metrics ($|V \cap M_{DS}|$), and average defect percentage in $|V \cap M_{DS}|$. 

For the experiments, we used implementations available on Github\cite{github}. The \textit{Node2Vec} implementation is the one released by the authors\cite{node2vec}. The \textit{LINE} implementation is part of the OpenNE toolkit\cite{openne}. For all embedding methods we used an embedding dimesion of 32, and used the default parameters. Specifically, we used $p$ and $q$ (for \textit{Node2Vec}) to be 1.

To evaluate the performance of our methods and the baseline methods, we used two commonly used performance metrics, Area Under the ROC Curve (AUC) and the F1-Score. F1-Score is defined by 

$$F_{1} = 2 \cdot \frac{precision \cdot recall}{precision + recall}$$

The results of the different setups were compared statistically using the Wilcoxon signed-rank test\cite{demsar_statistical_2006}. We performed experiments and compared the results on the following scenarios:

\subsubsection{Static Code Metrics (Baseline)}
In this setup we simply trained a classifier only on the static code metrics (of the old version) and tested on the new version. This represents a baseline since most defect prediction studies rely on these features.

\subsubsection{Embedding with No Alignment (Baseline)}
In this setup we train a classifier on static features together with learned embeddings, but without performing the alignment process. This setup comes to show how not performing alignment usually degrades the performance of the model. This is due to the fact that the embedding algorithm is not constrained to learn the same semantic meaning of each of the embedding dimensions (although a close solution might happen by chance).

\subsubsection{Embedding Alignment with Random Anchor Selection (Baseline)}
As another baseline, we evaluated the performance of Linear Regression and Orthogonal Transformation on a randomly selected set of anchors. The number of anchors selected was also modified to get a broader result. 

\subsubsection{Embedding Alignment with KNN Anchor Selection}
For this scenario, we evaluated the performance of our KNN Anchor Selection algorithm. We experimented with different numbers of anchors and numbers of nearest neighbors to take into account. The experiments were done using both an Orthogonal Transformation and Linear Regression and on both embedding techniques.

\subsubsection{Embedding Alignment with Graph Neighbors Similarity Anchor Selection}
For this scenario, we evaluated the performance of our Graph Neighbors Similarity Anchor Selection algorithm. In this scenario, we modified the number of anchors selected. The experiments were done using both an Orthogonal Transformation and Linear Regression and on both embedding techniques.

\begin{table}[ht]
\begin{tabularx}{\linewidth}{||Y|Y|Y|| }
    \hline
    Project Name & Project Description & Version Pairs\\
    \hline\hline
    Apache Camel & Integration Framework & (1.0,1.2) (1.2,1.4) (1.4,1.6)\\
    \hline
    JEdit & Text Editor & (3.2,4.0) (4.0,4.1) (4.1,4.2) (4.2,4.3)\\
    \hline
    Apache Log4J & Logging Library & (1.0,1.1) (1.1,1.2)\\
    \hline
    Apache Lucene & Information Retrieval Library & (2.0,2.2) (2.2,2.4)\\
    \hline
    Apache POI & Microsoft Office processing library & (1.5,2.0) (2.0,2.5) (2.5 3.0)\\
    \hline
    Apache Synapse & Enterprise Service Bus & (1.0,1.1) (1.1,1.2)\\
    \hline
    Apache Velocity & Template Engine & (1.4,1.5) (1.5,1.6)\\
    \hline
    Apache Xalan & XSLT and XPath implementation  & (2.4,2.5) (2.5,2.6) (2.6,2.7)\\
    \hline
    Apache Xerces & XML Processing Library & (init,1.2) (1.2,1.3) (1.3,1.4)\\
    \hline
\end{tabularx}
\caption{Projects Analyzed In Our Experiment}
\label{table:3}
\end{table}

\begin{table}[ht]
\begin{tabularx}{\linewidth}{||Y|Y|Y|Y|Y|| }
\hline
Project Name & Average $|V|$ & Average $|E|$ & Average $|V \cap M_{DS}|$ & Average Defect Percentage \\ \hline
Apache Camel        & 1312      & 4856      & 664               & 19.7                      \\ \hline
JEdit        & 662        & 2449       & 340                & 18.7                      \\ \hline
Apache Log4J        & 225  & 626          & 133          & 51.03               \\ \hline
Apache Lucene       & 1069         & 5197  & 257          & 55.5               \\ \hline
Apache POI          & 814       & 3282       & 341                  & 50                     \\ \hline
Apache Synapse      & 414  & 1470  & 209                  & 23.6               \\ \hline
Apache Velocity     & 356  & 1311 & 209          & 59                        \\ \hline
Apache Xalan        & 979       & 4535       & 805                & 52.2                     \\ \hline
Apache Xerces       & 514          & 1940      & 336               & 35.4               \\ \hline

\end{tabularx}
\caption{Dataset Statistics}
\label{table:4}
\end{table}

\section{Experimental Results}\label{ExpRes}
During our experiments, many different techniques and setups were evaluated. We wish to provide a few points of view on the different results, so this section will be divided into a few subsections that each analyze a different aspect of the results.

\subsubsection{Best Results for Each Embedding Algorithm}

\begin{figure*}[t]
\centering
\includegraphics[width=\textwidth]{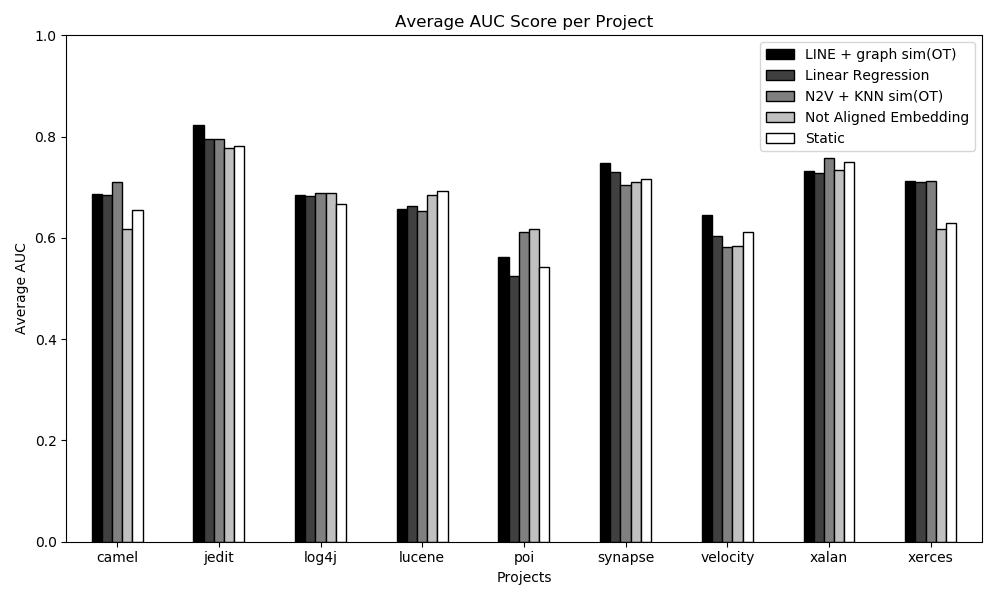}  
\caption{AUC results of our best methods versus three baseline models}
\label{fig:best}

\includegraphics[width=\textwidth]{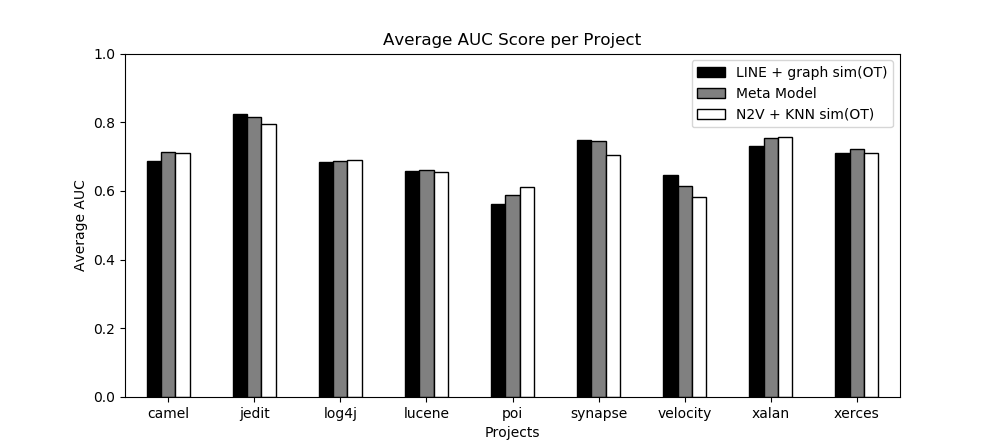}
\caption{AUC results of the joint model versus the individual models}
\label{fig:meta}
\end{figure*}

As described in an earlier section, we evaluated two embedding techniques, two anchor selection techniques, and two alignment techniques. In this section, we will provide the results of our best model for each of the embedding techniques. For \textit{LINE} embedding, we achieved the best results (in AUC terms) by using Orthogonal Transformation as the alignment and used the graph similarity anchor selection. These results were statistically significant with $p = 0.01$, compared to the static metrics model. For \textit{Node2vec}, we achieved the best results (in AUC terms) again using Orthogonal Transformation but using KNN anchor selection instead. The \textit{Node2vec} results were also statistically significant, with $p < 0.043$, compared to static metrics model. Figure \ref{fig:best} shows the AUC scores for the two methods just described and three baseline methods. The \textit{Static} method is simply a classifier with static code metrics. The \textit{Not Aligned Embedding} method is a classifier trained on both the static code metrics and the embeddings, but without aligning the embeddings. The \textit{Linear Regression} method uses the same embedding but uses Linear Regression as an embedding alignment mechanism. It is also interesting to note that for the Linear Regression mapping, we achieved the best result when we used all available code modules in both versions as our anchors. This result will be discussed in a later section. The results show that using CDN data provides an improvement in AUC performance. We also measured the F1-score of the different methods we used. In most cases, we got very similar results to the baseline, up to $\pm0.5\%$ difference. The results also show that Orthogonal Transformation alignment performs better than Linear Regression, although these results were not statistically significant. The not aligned model generally provides the worst results, but in some cases it outperforms. This seems like an anomaly, as was discussed earlier.

Another interesting phenomenon we can see from the results in Figure \ref{fig:best} is that the embedding techniques we used are somewhat complementary. In some projects, the results are similar, but there are projects for which one embedding technique is better than the other and vice versa. One possible explanation can be that different embedding techniques (and parameters) extract different information, specifically local vs. global information. Both local and global information have been shown to be relevant to defect prediction \cite{zimmermann_predicting_2008-1}. This difference in performance led us to create a meta-model using the two models we described in this section, and we will discuss it and its results in the next section.

\subsubsection{LINE and Node2vec joint model}
As we described in the previous section, the results of the \textit{LINE} model and the \textit{Node2vec} model complement each other. To utilize this result, we built a Logistic Regression model on top of the individual models. We extracted from each classifier the probability of a defect and used the Logistic Regression model to calculate a better probability estimate. On average, the model improves the individual model's results by about 1\%. The results are shown in Figure \ref{fig:meta}. A representative ROC curve for the meta-model versus the static metrics model is shown in Figure \ref{fig:roc}. We also checked the results for statistical significance and concluded that they are significant with $p < 0.002$, compared to the static metrics model. They were also statically significant when compared to the Linear Regression results, with $p < 0.001$.

\subsubsection{KNN Anchor Selection Analysis}
\begin{figure}[ht]
\centering
\begin{subfigure}{\linewidth}
\centering
\includegraphics[width=8cm]{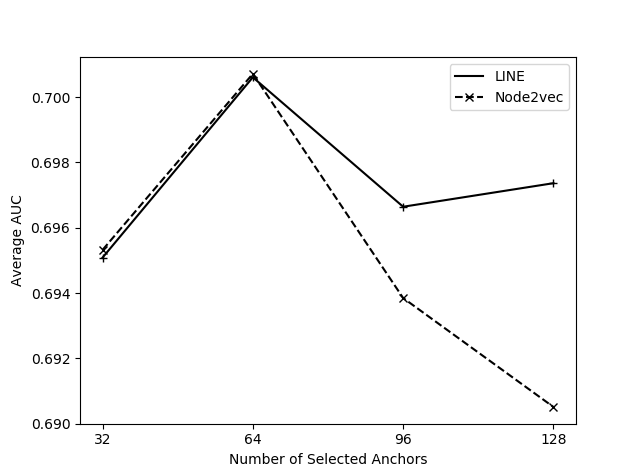}  
\caption{}
\label{fig:KNN_AUC}
\end{subfigure}
\begin{subfigure}{\linewidth}
\centering
\includegraphics[width=8cm]{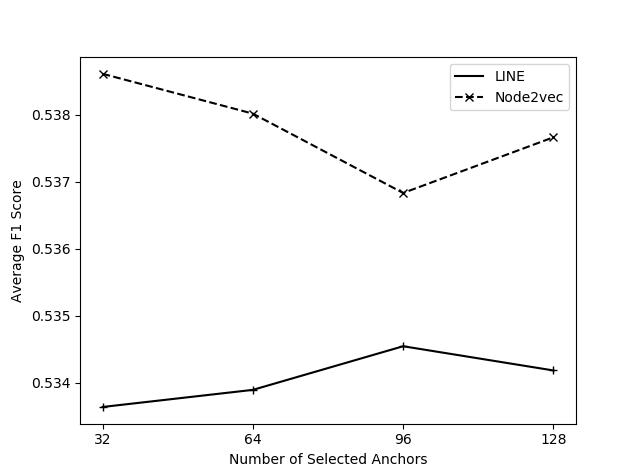}  
\caption{}
\label{fig:KNN_F1}
\end{subfigure}
\caption{KNN anchor selection method AUC (\ref{fig:KNN_AUC}) and F1 (\ref{fig:KNN_F1}) results for all embeddings  }
\label{fig:KNN}
\end{figure}

As described earlier, KNN anchor selection has two parameters: The number of anchors to select and $K$, which is the number of nearest neighbors for each candidate anchor to compare. From the experiments, it appeared that the number of anchors to select is the more significant parameter, especially when using the Orthogonal Transformation embedding alignment. Figure \ref{fig:KNN} shows how the number of anchors impacts the classification performance for all the embedding techniques. For these experiments, $K$ was fixed at 10. Increasing $K$ values gave us similar or worse results, in all experiments. It can be seen that using this setup, \textit{Node2vec} achieves the best performance. 

\subsubsection{Graph Similarity Anchor Selection Analysis}
\begin{figure}[ht]
\centering
\begin{subfigure}{\linewidth}
\centering
\includegraphics[width=8cm]{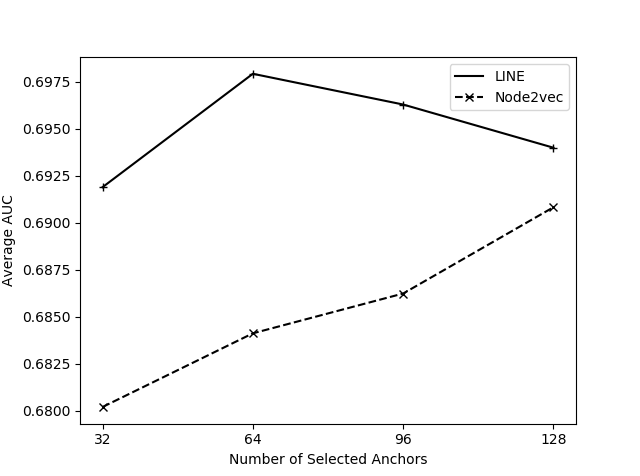}  
\caption{}
\label{fig:GS_AUC}
\end{subfigure}
\begin{subfigure}{\linewidth}
\centering
\includegraphics[width=8cm]{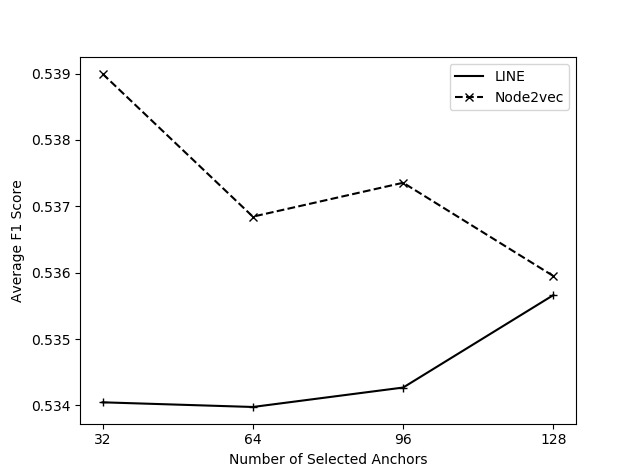}  
\caption{}
\label{fig:GS_F1}
\end{subfigure}
\caption{Graph similarity anchor selection method AUC (\ref{fig:GS_AUC}) and F1 (\ref{fig:GS_F1}) results for all embeddings  }
\label{fig:GraphSim}
\end{figure}

\begin{figure}[ht]
\includegraphics[width=\linewidth]{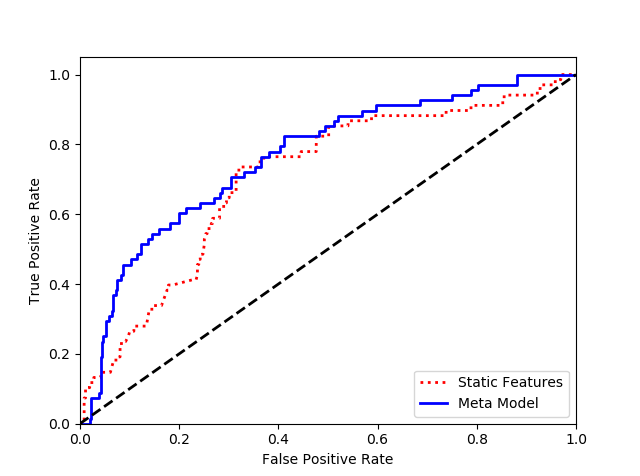}  
\caption{ROC curve for the Meta Model vs Static Metrics model on the (Xerces-init, Xerces-1.2) version pair}
\label{fig:roc}
\end{figure}

Similarly to the KNN analysis, we evaluated the impact of changing the number of anchors. The results of these experiments are shown in Figure \ref{fig:GraphSim}. In terms of AUC performance, it can be seen that \textit{LINE} achieves better results than the other two embedding techniques. This result is interesting because of how the \textit{LINE} embedding works. As described earlier, the main idea of \textit{LINE} is to model the neighborhoods of nodes. This anchor selection technique looks at precisely that. We try to find the nodes in both graphs that have the most similar neighborhoods. It seems reasonable that this technique would work best with \textit{LINE} and in general achieves the best AUC performance.

\begin{figure}[ht]
\includegraphics[width=\linewidth]{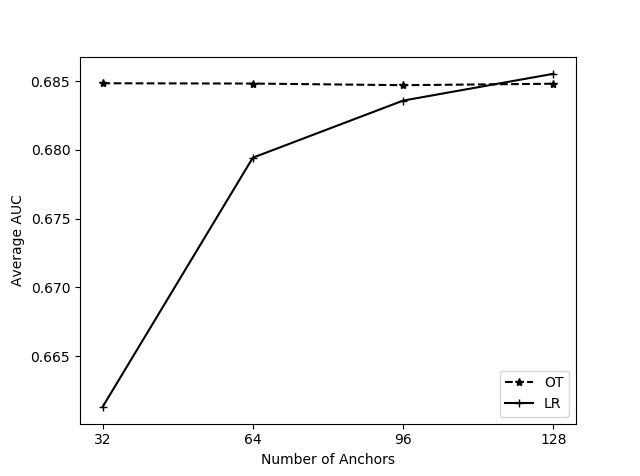}  
\caption{Average AUC for randomly selected anchors}
\label{fig:LR_OT}
\end{figure}

\subsubsection{Linear Regression and Orthogonal Transformation Comparison}
During this work, we used Linear Regression as a baseline method for the alignment process. Because Linear Regression learns a general linear transformation, some properties of the original space might be modified. For example, angles between vectors in the original domain can change after the transformation. This change in angles does not occur when an Orthogonal Transformation is used. Nevertheless, Linear Regression can still be a good approximation, and our experiments show this. 

In the section on the best model for each embedding, we presented the performance of Linear Regression as an alignment technique, while using the \textit{LINE} embedding. The results with the \textit{Node2vec} embedding were very similar and slightly worse. It can be seen that the performance of Linear Regression is usually worse or similar to that of the Orthogonal Transformation. These results shows that Linear Regression distorts some of the information available in the embedding. It appears that Orthogonal Transformations preserves more of the information in the embedding, hence achieving better results.

One more experiment we performed is to compare both alignment techniques in a random anchor selection setup. We performed 30 experiments for every anchor count and sampled randomly from the set of software modules that exist in both versions. Figure \ref{fig:LR_OT} shows the results of this experiment. For this experiment we used the \textit{Node2vec} embedding. The results show an interesting phenomenon. When we increase the number of anchors, Linear Regression achieves better results. On the other hand, randomly selecting anchors when using an Orthogonal Transformation does not achieve great results (less than with our selection techniques). It seems that Linear Regression achieves better alignment as more data points are added.

\section{Discussion}
The results presented in the previous section show an improvement over the basic model, which is based on static code metrics. From analyzing the results, one can see that the improvement is not uniform. Some projects exhibit better performance and some worse. As we discussed earlier, the different embedding techniques appeared to be complementary, and when these techniques were combined in a joint model, the overall performance improved. It still seems that there is more room for improvement because in some cases, the individual models beat the meta-model's performance. Analyzing these phenomenons is something we are looking at for future work.

As we mentioned before, we achieved the best results for different embedding techniques using different anchor selection techniques. This is an interesting and somewhat surprising result. This means that the notion of similarity and closeness in different embeddings is different. Because of this difference, it appears that there is a connection between the embedding algorithm's closeness notion and the anchor selection technique that works best and how it selects the best candidates. For this reason, a future direction will be to explore new anchor selection methods and match similar embedding techniques.

\section{Threats to Validity}
Our work might suffer from threats to its validity. We discuss them briefly.

\subsection{Threats to Internal Validity}
In our work, we measured the performance of our techniques using the widely used AUC and F1-Score. First, there are other performance measures which we did not use and might have different performance than we observed. For this reason, our results might not be relevant to some applications. Second, we observed a slight decrease in F1-Score, and believe it to be negligible. In some scenarios this might not be negligible, and so our conclusions might be mistaken. Nevertheless, this difference was not statistically significant.

\subsection{Threats to External Validity}
We have used the dataset collected by Jureczko et al. in our evaluations. This is a widely used dataset and contains several applications from different domains and of different sizes. There is a possibility that on a different dataset, we will achieve different results. This dataset dates back a few years and might not reflect changes in the Software Development Community. Another issue is that we analyzed Java applications, and on a different programming language, the results might be different.

\section{Conclusion And Future Work}
In this work, we aimed at improving the results of CVDP using Class Dependency Network data. For this purpose, we developed a framework for embedding and aligning CDN data and used its results as inputs for a classifier. We also suggested two anchor selection techniques and used them in different embedding and alignment setups. We performed extensive experiments using two embedding techniques on a publicly available dataset. Our results show that:
\begin{enumerate}
    \item As previously shown, CDN data is beneficial for defect prediction. We showed that this is true for the CVDP scenario as well.
    \item We developed a framework for the generation and alignment of embeddings of CDNs across versions.
    \item We performed multiple experiments and showed that our framework achieves statistically better performance than the state-of-the-art baseline.
\end{enumerate}

For future work, we are considering the following directions:
\begin{enumerate}
    \item Experiment with more embedding techniques and different parameter settings. These settings can provide more local vs. global information into the embedding process. 
    \item Experiment with new datasets, in order to provide a broader performance measure.
    \item Try new approaches for CDN embedding that take into account labels and weights.
    \item Try to use multiple old versions (MOV) in the learning process, instead of just a single old version (SOV).
\end{enumerate}

\bibliography{refs}

\begin{thebibliography}{}

\bibitem[git, ]{github}
Github.
\newblock \url{https://github.com/}.

\bibitem[noa, ]{noauthor_java_nodate}
Java {Software} {\textbar} {Oracle}.
\newblock \url{https://www.oracle.com/java/}.

\bibitem[nod, ]{node2vec}
Node2vec.
\newblock \url{https://github.com/aditya-grover/node2vec}.

\bibitem[ope, ]{openne}
{OpenNE} - {An} {Open}-{Source} {Package} for {Network} {Embedding} ({NE}).
\newblock \url{https://github.com/thunlp/OpenNE}.

\bibitem[Amasaki, 2018]{amasaki_cross-version_2018}
Amasaki, S. (2018).
\newblock Cross-{Version} {Defect} {Prediction} {Using} {Cross}-{Project}
  {Defect} {Prediction} {Approaches}: {Does} {It} {Work}?
\newblock In {\em Proceedings of the 14th {International} {Conference} on
  {Predictive} {Models} and {Data} {Analytics} in {Software} {Engineering}},
  {PROMISE}'18, pages 32--41, New York, NY, USA. ACM.

\bibitem[Bansiya and Davis, 2002]{bansiya_hierarchical_2002}
Bansiya, J. and Davis, C.~G. (2002).
\newblock A hierarchical model for object-oriented design quality assessment.
\newblock {\em IEEE Transactions on Software Engineering}, 28(1):4--17.

\bibitem[Bennin et~al., 2017]{bennin_significant_2017}
Bennin, K.~E., Keung, J., Monden, A., Phannachitta, P., and Mensah, S. (2017).
\newblock The {Significant} {Effects} of {Data} {Sampling} {Approaches} on
  {Software} {Defect} {Prioritization} and {Classification}.
\newblock In {\em Proceedings of the 11th {ACM}/{IEEE} {International}
  {Symposium} on {Empirical} {Software} {Engineering} and {Measurement}},
  {ESEM} '17, pages 364--373, Piscataway, NJ, USA. IEEE Press.

\bibitem[Bennin et~al., 2016]{bennin_empirical_2016}
Bennin, K.~E., Toda, K., Kamei, Y., Keung, J., Monden, A., and Ubayashi, N.
  (2016).
\newblock Empirical {Evaluation} of {Cross}-{Release} {Effort}-{Aware} {Defect}
  {Prediction} {Models}.
\newblock In {\em 2016 {IEEE} {International} {Conference} on {Software}
  {Quality}, {Reliability} and {Security} ({QRS})}, pages 214--221.

\bibitem[Breiman, 2001]{breiman_random_2001}
Breiman, L. (2001).
\newblock Random {Forests}.
\newblock {\em Machine Learning}, 45(1):5--32.

\bibitem[Chidamber and Kemerer, 1994]{chidamber_metrics_1994}
Chidamber, S.~R. and Kemerer, C.~F. (1994).
\newblock A metrics suite for object oriented design.
\newblock {\em IEEE Transactions on Software Engineering}, 20(6):476--493.

\bibitem[D'Ambros et~al., 2010]{dambros_extensive_2010}
D'Ambros, M., Lanza, M., and Robbes, R. (2010).
\newblock An extensive comparison of bug prediction approaches.
\newblock In {\em 2010 7th {IEEE} {Working} {Conference} on {Mining} {Software}
  {Repositories} ({MSR} 2010)}, pages 31--41.

\bibitem[Demšar, 2006]{demsar_statistical_2006}
Demšar, J. (2006).
\newblock Statistical {Comparisons} of {Classifiers} over {Multiple} {Data}
  {Sets}.
\newblock {\em J. Mach. Learn. Res.}, 7:1--30.

\bibitem[Fan et~al., 2019]{fan_software_2019}
Fan, G., Diao, X., Yu, H., Yang, K., and Chen, L. (2019).
\newblock Software {Defect} {Prediction} via {Attention}-{Based} {Recurrent}
  {Neural} {Network}.
\newblock {\em Scientific Programming}, 2019:14.

\bibitem[Fenton and Bieman, 2014]{fenton_software_2014}
Fenton, N. and Bieman, J. (2014).
\newblock {\em Software metrics: a rigorous and practical approach}.
\newblock CRC press.

\bibitem[Goyal and Ferrara, 2017]{goyal_graph_2017}
Goyal, P. and Ferrara, E. (2017).
\newblock Graph {Embedding} {Techniques}, {Applications}, and {Performance}:
  {A} {Survey}.
\newblock {\em CoRR}, abs/1705.02801.

\bibitem[Grover and Leskovec, 2016]{grover_node2vec_2016}
Grover, A. and Leskovec, J. (2016).
\newblock node2vec: {Scalable} {Feature} {Learning} for {Networks}.
\newblock In {\em Proceedings of the 22nd {ACM} {SIGKDD} {International}
  {Conference} on {Knowledge} {Discovery} and {Data} {Mining}}.

\bibitem[Hall et~al., 2012]{hall_systematic_2012}
Hall, T., Beecham, S., Bowes, D., Gray, D., and Counsell, S. (2012).
\newblock A {Systematic} {Literature} {Review} on {Fault} {Prediction}
  {Performance} in {Software} {Engineering}.
\newblock {\em IEEE Transactions on Software Engineering}, 38(6):1276--1304.

\bibitem[Henderson-Sellers, 1996]{henderson-sellers_object-oriented_1996}
Henderson-Sellers, B. (1996).
\newblock {\em Object-oriented {Metrics}: {Measures} of {Complexity}}.
\newblock Prentice-Hall, Inc., Upper Saddle River, NJ, USA.

\bibitem[Jureczko and Madeyski, 2010]{jureczko_towards_2010}
Jureczko, M. and Madeyski, L. (2010).
\newblock Towards {Identifying} {Software} {Project} {Clusters} with {Regard}
  to {Defect} {Prediction}.
\newblock In {\em Proceedings of the 6th {International} {Conference} on
  {Predictive} {Models} in {Software} {Engineering}}, {PROMISE} '10, pages
  9:1--9:10, New York, NY, USA. ACM.

\bibitem[Kamei and Shihab, 2016]{kamei_defect_2016-1}
Kamei, Y. and Shihab, E. (2016).
\newblock Defect {Prediction}: {Accomplishments} and {Future} {Challenges}.
\newblock In {\em 2016 {IEEE} 23rd {International} {Conference} on {Software}
  {Analysis}, {Evolution}, and {Reengineering} ({SANER})}, volume~5, pages
  33--45.

\bibitem[Ma et~al., 2016]{ma_empirical_2016}
Ma, W., Chen, L., Yang, Y., Zhou, Y., and Xu, B. (2016).
\newblock Empirical analysis of network measures for effort-aware
  fault-proneness prediction.
\newblock {\em Information and Software Technology}, 69:50 -- 70.

\bibitem[Martin, 1994]{martin_oo_1994}
Martin, R. (1994).
\newblock {OO} design quality metrics.

\bibitem[McCabe, 1976]{mccabe_complexity_1976}
McCabe, T.~J. (1976).
\newblock A {Complexity} {Measure}.
\newblock {\em IEEE Transactions on Software Engineering}, SE-2(4):308--320.

\bibitem[Mikolov et~al., 2013]{mikolov_efficient_2013}
Mikolov, T., Chen, K., Corrado, G., and Dean, J. (2013).
\newblock Efficient {Estimation} of {Word} {Representations} in {Vector}
  {Space}.
\newblock {\em CoRR}, abs/1301.3781.

\bibitem[Nam, 2014]{nam_survey_2014}
Nam, J. (2014).
\newblock Survey on {Software} {Defect} {Prediction}.
\newblock Tech {Report}, The Hong Kong University of Science and Technology,
  Department of Compter Science and Engineerning.

\bibitem[Premraj and Herzig, 2011]{premraj_network_2011}
Premraj, R. and Herzig, K. (2011).
\newblock Network {Versus} {Code} {Metrics} to {Predict} {Defects}: {A}
  {Replication} {Study}.
\newblock In {\em 2011 {International} {Symposium} on {Empirical} {Software}
  {Engineering} and {Measurement}}, pages 215--224.

\bibitem[Qu et~al., 2018]{qu_node2defect_2018}
Qu, Y., Liu, T., Chi, J., Jin, Y., Cui, D., He, A., and Zheng, Q. (2018).
\newblock Node2defect: {Using} {Network} {Embedding} to {Improve} {Software}
  {Defect} {Prediction}.
\newblock In {\em Proceedings of the 33rd {ACM}/{IEEE} {International}
  {Conference} on {Automated} {Software} {Engineering}}, {ASE} 2018, pages
  844--849, New York, NY, USA. ACM.

\bibitem[Qu et~al., 2019]{qu_using_2019}
Qu, Y., Zheng, Q., Chi, J., Jin, Y., He, A., Cui, D., Zhang, H., and Liu, T.
  (2019).
\newblock Using {K}-core {Decomposition} on {Class} {Dependency} {Networks} to
  {Improve} {Bug} {Prediction} {Model}'s {Practical} {Performance}.
\newblock {\em IEEE Transactions on Software Engineering}, pages 1--1.

\bibitem[Schönemann, 1966]{schonemann_generalized_1966}
Schönemann, P.~H. (1966).
\newblock A generalized solution of the orthogonal procrustes problem.
\newblock {\em Psychometrika}, 31(1):1--10.

\bibitem[Shukla et~al., 2018]{shukla_multi-objective_2018}
Shukla, S., Radhakrishnan, T., Muthukumaran, K., and Neti, L. B.~M. (2018).
\newblock Multi-objective cross-version defect prediction.
\newblock {\em Soft Computing}, 22(6):1959--1980.

\bibitem[Smith et~al., 2017]{smith_offline_2017}
Smith, S.~L., Turban, D. H.~P., Hamblin, S., and Hammerla, N.~Y. (2017).
\newblock Offline bilingual word vectors, orthogonal transformations and the
  inverted softmax.
\newblock {\em CoRR}, abs/1702.03859.

\bibitem[Subelj and Bajec, 2011]{subelj_community_2011}
Subelj, L. and Bajec, M. (2011).
\newblock Community structure of complex software systems: {Analysis} and
  applications.
\newblock {\em CoRR}, abs/1105.4276.

\bibitem[Tang et~al., 2015]{tang_line_2015}
Tang, J., Qu, M., Wang, M., Zhang, M., Yan, J., and Mei, Q. (2015).
\newblock {LINE}: {Large}-scale {Information} {Network} {Embedding}.
\newblock {\em CoRR}, abs/1503.03578.

\bibitem[Tang et~al., 1999]{tang_empirical_1999}
Tang, M.-H., Kao, M.-H., and Chen, M.-H. (1999).
\newblock An empirical study on object-oriented metrics.
\newblock In {\em Proceedings {Sixth} {International} {Software} {Metrics}
  {Symposium} ({Cat}. {No}.{PR}00403)}, pages 242--249.

\bibitem[Xing et~al., 2015]{xing_normalized_2015}
Xing, C., Wang, D., Liu, C., and Lin, Y. (2015).
\newblock Normalized {Word} {Embedding} and {Orthogonal} {Transform} for
  {Bilingual} {Word} {Translation}.
\newblock In {\em Proceedings of the 2015 {Conference} of the {North}
  {American} {Chapter} of the {Association} for {Computational} {Linguistics}:
  {Human} {Language} {Technologies}}, pages 1006--1011, Denver, Colorado.
  Association for Computational Linguistics.

\bibitem[Xu et~al., 2019]{XU201959}
Xu, Z., Li, S., Luo, X., Liu, J., Zhang, T., Tang, Y., Xu, J., Yuan, P., and
  Keung, J. (2019).
\newblock {TSTSS}: {A} two-stage training subset selection framework for cross
  version defect prediction.
\newblock {\em Journal of Systems and Software}, 154:59--78.

\bibitem[Xu et~al., 2018a]{xu_cross_2018}
Xu, Z., Li, S., Tang, Y., Luo, X., Zhang, T., Liu, J., and Xu, J. (2018a).
\newblock Cross {Version} {Defect} {Prediction} with {Representative} {Data}
  via {Sparse} {Subset} {Selection}.
\newblock In {\em Proceedings of the 26th {Conference} on {Program}
  {Comprehension}}, {ICPC} '18, pages 132--143, New York, NY, USA. ACM.

\bibitem[Xu et~al., 2018b]{xu_cross-version_2018}
Xu, Z., Liu, J., Luo, X., and Zhang, T. (2018b).
\newblock Cross-version defect prediction via hybrid active learning with
  kernel principal component analysis.
\newblock In {\em 2018 {IEEE} 25th {International} {Conference} on {Software}
  {Analysis}, {Evolution} and {Reengineering} ({SANER})}, pages 209--220.

\bibitem[Yang and Wen, 2018]{yang_ridge_2018}
Yang, X. and Wen, W. (2018).
\newblock Ridge and {Lasso} {Regression} {Models} for {Cross}-{Version}
  {Defect} {Prediction}.
\newblock {\em IEEE Transactions on Reliability}, 67(3):885--896.

\bibitem[Zimmermann and Nagappan, 2008]{zimmermann_predicting_2008-1}
Zimmermann, T. and Nagappan, N. (2008).
\newblock Predicting defects using network analysis on dependency graphs.
\newblock In {\em 2008 {ACM}/{IEEE} 30th {International} {Conference} on
  {Software} {Engineering}}, pages 531--540.

\bibitem[Zimmermann et~al., 2007]{zimmermann_predicting_2007}
Zimmermann, T., Premraj, R., and Zeller, A. (2007).
\newblock Predicting {Defects} for {Eclipse}.
\newblock In {\em Proceedings of the {Third} {International} {Workshop} on
  {Predictor} {Models} in {Software} {Engineering}}, {PROMISE} '07, pages 9--,
  Washington, DC, USA. IEEE Computer Society.

\end{thebibliography}
\end{document}